\documentclass[namedreferences,hyperref,optionalrh]{spr-sola}
\usepackage{graphicx}        
\usepackage{color}           
\usepackage{amsmath,amssymb}

\def\ssn{S}

\def\devLQ{\mbox{dev}_{\scriptsize LQ}}

\newcommand{\mo}[1]{{\textcolor{black}{#1}}}
\newcommand{\rev}[1]{{\textcolor{black}{#1}}}
\usepackage{fix-cm}

\usepackage{natbib}
\setcitestyle{aysep={}}

\chardef\us=`\_
\newcommand{\mm}[1]{{\textcolor{black}{#1}}}

\begin{document}

\begin{frontmatter}
\title{Effect of Nonlinear Surface Inflows into Activity Belts on Solar Cycle Modulation}

\author[addressref={aff1,aff2}, corref,email={mtalafha901@gmail.com}]{\inits{M. H.}\fnm{Mohammed H.}~\snm{Talafha}\orcid{0000-0003-4671-1759}}

\author[addressref=aff3,email={k.petrovay@astro.elte.hu}]{\inits{K.}\fnm{Kristóf }~\snm{Petrovay}\orcid{0000-0002-1714-2027}}

\author[addressref=aff1,email={opitz.andrea@wigner.hu}]{\inits{A.}\fnm{Andrea}~\snm{Opitz}\orcid{0000-0003-0845-0201}}

\address[id=aff1]{HUN-REN Wigner Research Centre for Physics, Department of Space Physics and Space Technology, Budapest, Hungary}

\address[id=aff2]{\rev{Research Institute of Science and Engineering, University of Sharjah, Sharjah, UAE}}

\address[id=aff3]{ELTE Eötvös Loránd University, Institute of Geography and Earth Sciences, Department of Astronomy, Budapest, Hungary}

\runningauthor{Talafha et al.}
\runningtitle{\mm{E}ffect of \mm{S}urface \mm{I}nflows}

\begin{abstract}
Converging flows are visible around {bipolar magnetic regions (BMRs)} on the solar surface, according to observations. Average flows are created by these inflows combined, and the strength of these flows depends on the amount of flux present during the solar cycle. In models of the solar cycle, this average flow can be depicted as perturbations to the meridional flow. \mm{In this article}, we study the effects of introducing surface inflow to the surface flux transport models (SFT) as a possible nonlinear mechanism in the presence of {latitude} quenching for an inflow profile whose amplitude varies within a cycle depending on the magnetic activity. Using a grid based on one-dimensional SFT models, we methodically investigated the extent of nonlinearity caused by inflows, latitude quenching (LQ), and their combinations. The results show that including surface inflows in the model in the presence of both LQ and tilt quenching (TQ) produced a polar field within a \mm{$\pm$1$\sigma$} of an average cycle polar field \mm{($\sigma$ is the standard deviation)} and a correlation coefficient of 0.85. We confirm that including inflows produces a lower net contribution to the dipole moment (\mm{10\,--\,25\%}). Furthermore, the relative importance of LQ vs. inflows is inversely correlated with the dynamo effectivity range ($\lambda_{R}$). With no decay term, introducing inflows into the model resulted in a less significant net contribution to the dipole moment. Including inflows in the SFT model shows a possible nonlinear relationship between the surface inflows and the solar dipole moment, suggesting a potential nonlinear mechanism contributing to the saturation of the global dynamo. For lower $\lambda_R$ ($\lessapprox$ 10 $^\circ$), TQ always dominates LQ, and for higher $\lambda_R$ LQ  dominate. However, including inflows will make the domination a little bit earlier in case of having a decay term in the model.
\end{abstract}
\end{frontmatter}

\section{Introduction}
\label{sec:intro}

Solar dynamo models describe the evolution of solar cycles over time, with poloidal and toroidal fields as the main components in a Babcock-Leighton (BL) dynamo model. Poloidal fields wind up by differential rotation to form the toroidal fields of the next cycle, and the toroidal field serves as seeds for the poloidal field \citep{bhowmik2023physical}. Surface flux transport models (SFT) have been used to describe these large-scale photospheric magnetic fields' evolution by assuming that the surface field is nearly radial. Its evolution can be represented by the radial component of the magnetohydrodynamics (MHD) induction equation, where advection is attributed to differential rotation and poleward meridional flow and diffusion to the mixing action of supergranular flows \citep{yeates2023surface}. 
The reversal of the polar field and solar dipole moment in the middle of the activity cycles originates from the systematic latitude-dependent tilt of the active regions (ARs) relative to the azimuthal direction (Joy's law) and can be reproduced by the SFT. In addition, \cite{schrijver2002missing} introduced a decay term to solve the problem of dipole moment drift. SFT models were used in literature for simulating and modelling the solar magnetic field \citep{cameron2012surface, Cameron+Schussler:turbdiff, Jiang+:1700a, Whitbread+:SFT};  in addition to solar cycle prediction \citep{jiang2018predictability, Iijima+:plateau, Jiang:nonlin, Petrovay:LRSP2}, a recent review of the model given by \cite{yeates2023surface}.


The generation of the toroidal magnetic field is linearly linked to the strength of the poloidal field during the minimum of the previous cycle  \citep{jiang2018predictability}. However, it is anticipated that the transformation of the toroidal field into the poloidal field through the BL mechanism is a nonlinear process, which plays a significant role in the amplified growth of amplitude and variability of the solar cycle. 

The tilt angle of the sunspot group plays a vital role in the formation of the poloidal field generated by the toroidal field through the BL mechanism. \mo{As a sunspot group develops across the solar surface, its tilt angle changes over time. Analysing its characteristics at various stages of evolution provides insight into the physical origins of tilt angles. According to the thin flux-tube model of BMR formation, the tilt is thought to result from the Coriolis force acting on the rising flux tube of the strong toroidal magnetic field originating at the base of the convection zone \citep{jha2020magnetic}. The scatter in the tilt angle arises from two factors, one of which is the generation mechanism of the tilt angle, potentially influenced by turbulent convection. The other factor contributing to tilt angle scatter is measurement errors, such as those arising from measurements conducted in unipolar regions. Studying tilt angles aids in uncovering the mechanisms behind the randomness of the solar cycle and provides insights into flux tube emergence \citep{jiao2021sunspot}. } 

The thin-flux-tube approximation theory posits that as the field strength rises, there is a decrease in tilts, which implies a nonlinear feedback mechanism that affects the efficiency of the BL mechanism \citep{d1993theoretical}. 
\cite{Dasi-Espuig+} conducted a study that revealed a noteworthy association between the average tilt of bipolar active regions and the amplitude of the solar cycle. This correlation, commonly referred to as tilt-quenching (TQ), has been widely incorporated into solar dynamo models. Importantly, they found that the product of the mean tilt and cycle amplitude was a reliable predictor of the amplitude of the subsequent solar cycle. This finding holds significance in the context of understanding and forecasting solar activity dynamics. This correlation was confirmed by \cite{jiao2021sunspot} through a comprehensive examination of the existing methodologies employed to estimate tilt angles using data from the Kodaikanal and Mount Wilson observatories. In addition, the tilt angles from the Debrecen photoheliographic dataset were incorporated into the analysis. Meticulous analysis revealed a significant inverse correlation between the tilt angles and the strength of the solar cycle. Using line-of-sight magnetograms obtained from the Michelson Doppler Imager (MDI) aboard the Solar and Heliospheric Observatory (SOHO) covering the period 1996-2011, as well as the Helioseismic and Magnetic Imager (HMI) aboard the Solar Dynamic Observatory (SDO) from 2010 to 2018. \cite{jha2020magnetic} investigated the tilt characteristics of Bipolar Magnetic Regions (BMRs) within a solar cycle and demonstrated that the tilt of BMRs exhibits a nonmonotonic relationship with the field strength of the region. Specifically, the tilt angle increased in the small magnetic field regime, whereas it decreased in the large magnetic field regime.


The efficiency of poloidal field generation by BMRs diminishes when they emerge at higher latitudes, mainly because of the limited cross-equatorial cancellation \citep{jiang2014magnetic}. \rev{Cross-equatorial transport is indispensable for the buildup of the new poloidal field in the Babcock-Leighton scenario. Without cross-equatorial transport, there would be simply no polar reversal, and no new cycle would follow \citep{Petrovay:LRSP2}. For clarification, the cross-equatorial transport discussed \mm{in this article} is not related to the positions or proper motions of ARs: it is a diffusive transport of the large-scale mean-field, for which direct evidence would be hard to find. The overall agreement of the evolution of the large-scale solar magnetic field with SFT model predictions is, however, strong indirect evidence for cross-equatorial transport \citep{yeates2023surface}. } However, when a BMR emerges in proximity to the equator, the process of cancelling out the leading polarities with a flux of opposite polarity from the other hemisphere is facilitated. \cite{Jiang:nonlin} directed attention towards "latitude quenching" (LQ), as this nonlinear modulation mechanism is {coined by} \cite{Petrovay:LRSP2}. The LQ was derived from an empirical observation made through extensive analysis of a long sunspot record, revealing a correlation between the average latitude of the active region appearance during a specific phase of the solar cycle and the amplitude of the cycle \rev{\citep{Jiang+:1700a}}. The presence of active regions at higher latitudes hampers the diffusion of the leading flux across the equator, consequently reducing the contribution of the trailing flux to polar fields. Consequently, the observed negative correlation between the average latitude of active region emergence and the cycle amplitude signifies the presence of a negative feedback effect. \cite{Karak2020} argued that LQ can effectively govern the growth of the magnetic field, particularly when the dynamo operates within a moderately supercritical regime, by incorporating this concept into a 3D Babcock–Leighton dynamo model using a straightforward latitude-dependent threshold for BMR eruption.

\cite{paper2} conducted a comprehensive investigation to assess the influence of the {two potential nonlinearities, TQ and LQ, on} the source term in an SFT model systematic simulation. Both TQ and LQ serve as plausible saturation mechanisms that explain the absence of the runaway exponential growth in the solar dynamo. The study revealed that the relative impact of {TQ} versus {LQ} on the axial dipole at the end of the solar cycle is primarily determined by the dynamo effectivity range $\lambda_{R}$, {defined as a universal function of $\Delta_{u}/\eta$, where $\Delta_{u}$ is the divergence of the meridional flow at the equator and $\eta$ represents the diffusivity; Eqs. (26) and (27) in \cite{Petrovay+:algebraic1}}. Specifically, when $\lambda_{R}$ is small, LQ exhibits a greater reduction in the end-of-cycle axial dipole than TQ. Conversely, for large values of $\lambda_{R}$, TQ is more influential than LQ in diminishing the end-of-cycle dipole.

Observations on the solar surface reveal the presence of converging flows surrounding BMRs \citep{gizon2001probing, gonzalez2008subsurface}. These inflows collectively generate average flows around the activity belt, and their intensity is contingent on the level of flux within the solar cycle, as demonstrated by \cite{jiang2010effect} and \cite{cameron2012strengths}. The average poleward meridional flow experiences a modest change of approximately 25$\%$ between the maximum and minimum phases of the solar cycle, or from one cycle to another \citep{komm1993meridional, hathaway2010variations}. This variation is attributed, at least in part, to surface inflows directed towards large active regions \citep{gizon2004helioseismology, cameron2010changes, jiang2010effect}. These inflows can reach a substantial fraction of the mean axisymmetric poleward meridional flow at mid-latitudes, but their spatial extent remains confined to the belts where the active regions are located. The magnetic influence on surface flows can induce variations in the transportation of magnetic flux towards the pole, consequently affecting the dynamo efficiency of decaying bipolar active regions. This alteration leads to a reduction in the cross-equatorial cancellations of BMRs and suppresses the effectiveness of the Babcock–Leighton process. In the case of a strong solar cycle, this effect becomes more pronounced and imparts a stabilising influence on the dynamo, as demonstrated by \cite{martin2017inflows} and \cite{Nagy+:inflow}.


This study aims to investigate how surface inflows towards active-region belts affect solar cycle modulation as a quenching mechanism for the solar dynamo. In particular, we employ the model formulation introduced by \cite{jiang2010effect} and its subsequent modifications in \cite{Nagy+:inflow} by imposing axisymmetric bands of converging latitudinal flow as perturbations of meridional flow. \rev{The novelty in the current paper is the addition of the meridional inflow as opposed to either TQ or LQ .} 

The remainder of this paper is organized as follows. In \mm{Section}~\ref{sect:SFT}, we provide a detailed description of the SFT model used in this study and describe the surface inflow profile that we use. \mm{The results} of our solar cycle simulations are then presented in \mm{Section}~\ref{sect:SFTresults}, regarding the modified surface inflows profile \citep{Nagy+:inflow}. The discussion is given in \mm{Section}~\ref{sect:dis}, and the paper is concluded in \mm{Section}~\ref{sect:concl}.

\section{Modulation of \mm{S}urface \mm{I}nflows in a \mm{S}urface \mm{F}lux \mm{T}ransport \mm{M}odel}
\label{sect:SFT}
To investigate how surface inflows affect one-dimensional (1D) SFT models with the {LQ} mechanism, we used the results of \cite{paper1} (hereafter Paper~1) as the optimised combinations of model parameters and \cite{paper2} (hereafter Paper~2) as the parameter values for LQ.

As we modelled the "average" solar cycles in papers 1 and 2, the source function is a smooth distribution that can be regarded as an ensemble average. This allowed us to simulate the probability distribution of the emergence of leading and trailing polarity on the solar surface.
Because this source is axially symmetric, our entire SFT model can be simplified to one dimension, as shown in \mm{Equation}~\ref{eq:transp}.
\begin{eqnarray}
 \frac{\partial B}{\partial t} &=& 
 \frac{1}{R\cos{\lambda}}\frac{\partial}{\partial \lambda}(B\,u\,\cos{\lambda})
 \nonumber \\
 &&+\frac{\eta}{R^2\cos{\lambda}}
 \frac{\partial}{\partial \lambda}\left(\cos{\lambda}\frac{\partial B}{\partial \lambda}\right)
 -\frac{B}{\tau} + s(\lambda,t)
\label{eq:transp}
\end{eqnarray}
Where $u$ is the meridional flow speed, $\tau$ is the timescale of decay due to radial diffusion, $R$ is the solar radius, and $\eta$ is the supergranular diffusivity. \rev{The meridional flow profile used \mm{in this work} is within the observational constraints. In principle, dynamical models of meridional circulation might also be used; however, these also involve numerous ad hoc assumptions. For example, the recent numerical simulations by \cite{finley2024well} apply an arbitrary rescaling of the Nusselt number, given the convective conundrum, and they neglect the magnetic feedback. For our study, a dynamic model would be an unnecessary complication that would not make the flow pattern more unique.}

The source term $s$ represents flux emergence by a general form of ring pairs with opposite magnetic polarity, as described by \cite{paper2}; \mm{in this work}, we only considered the LQ effects (i.e., the $b_{joy}$ parameter is equal to zero). 

\rev{As explained in detail in the recent review by \cite{yeates2023surface}, the SFT equation is the radial component of the magnetic induction equation at the surface where the radial diffusion term is either neglected or represented in a simplified way (decay term with parameter in $\tau$ our case). A more realistic form for this term was given by \cite{baumann2004evolution}. According to one widely held view, the strong downwards-directed pumping in the shallow layers may make this
contribution insignificant (see \cite{ossendrijver2002magnetoconvection}; \cite{kapyla2006magnetoconvection}, and references therein). Apart from this simplification, the SFT equation is exact. Radial flows have no direct relevance to the redistribution of the radial magnetic field on the solar surface.}

Surface inflows were added as perturbations to the meridional flow $\Delta v(\lambda, t)$ as the axisymmetric parametric form introduced by \citet{jiang2010effect}, but their amplitude varies over time within a cycle depending on the magnetic activity \citep{Nagy+:inflow}. \rev{In a full non-axisymmetric model, the inflows should be centred on individual active regions. As, however, our study focuses on solar cycle modulation, which is determined by the axial dipole moment built up by the end of the cycle, it is sufficient \mm{in this work} to consider the azimuthally averaged SFT equation. The assumed latitudinal inflows result from the azimuthal averaging of the actual radial inflows towards ARs; hence, their amplitude is expected to scale with the number of ARs present and their time scale is comparable to the cycle length. Just as the time scale of the source term s is the cycle time rather than the lifetime of individual ARs. For the specific form (Equation~\ref{flow_v}) used for the scaling and the value of the amplitude parameter $v_{00}$, we follow \cite{Nagy+:inflow}}
{
\begin{equation}
\label{flow_pert1}
    \Delta v(\lambda, t) =
\left\{
 \begin{array}{ll}
 -v_{0}\sin(\frac{\lambda-\lambda_{0}(t)}{\delta\lambda})  
     & \mbox{if } -\pi < \frac{\lambda-\lambda_{0}(t)}{\delta\lambda} < \pi \\
 0 & \mbox{otherwise } 
 \end{array}
\right.  
\end{equation}}

Here, the velocity amplitude of the bands $v_{0}(t)$ is calculated as follows:
\begin{equation}
\label{flow_v}
    v_{0}(t) = v_{00} \cdot
 \arctan\left(\frac{S_{n}}{F_{00}}\right)
\end{equation}

$S_n$ is the scaled cycle amplitude and $F_{00}=4.99\times 10^{-21}$, is an arbitrary normalisation parameter, the initial amplitude of the inflows, $v_{00} = -500$ and $-750$ \mm{cm s$^{-1}$}. The dependence of the arctangent in \mm{Equation}~\ref{flow_v} prohibits unreasonably high inflow speeds for rogue BMR emergencies \citep{nagy2017effect}.
In these bands, $\delta\lambda$ is the width of the bands, and $\lambda_0$ is its central latitude. We obtain another band in the opposite hemisphere by substituting $\lambda_0$ $\rightarrow$ $-\lambda_0$. For sufficiently small central latitudes, the two bands can overlap, and the corresponding velocities are added. The temporal dependency of $\lambda_0(t;i)$, \mm{Equation}~\ref{eq:lambdan}, represents the equatorward travel of the bands during the solar cycle. The width of the bands $\delta\lambda$ follow the standard deviation in \mm{Equation}~\ref{eq:fwhm} 

\begin{equation}
\label{eq:lambdan}
\lambda_0(t;i) [^\circ]  =  [ 26.4 - 34.2 (t/P) + 16.1(t/P)^2 ] 
 (\lambda_{i}/14.6)
\end{equation}

\begin{equation}
\label{eq:fwhm}
\delta \lambda = [ 0.14+1.05(t/P)-0.78(t/P)^{2}]\lambda_{0} 
\end{equation}
where $P=11\,$year is the cycle period. 
\rev{From the point of view of our axisymmetric model, what matters is not the tilt angle by itself but the latitude difference between the leading and trailing polarities. This is not affected by differential rotation. Following usual practice, in our SFT model, the AR source is introduced instantaneously, and its further evolution is described by the SFT model.}

The amplitude of the inflows was scaled to depend on the cycle amplitude $A_n$, as ($A_0$-$A_n$)$\times 10$.  The constant $A_0$ was assigned a value of $0.001 \,exp({7 [yr]/\tau})$ to ensure that the resulting dipolar moments aligned reasonably well with the observed values (measured in Gauss) for an average cycle. Note that with this scale and $\tau$-dependence, defining $\ssn_n =\chi \,A_n s_1$ yields values of $\ssn_n$ comparable to the observed sunspot numbers (SSN), where $\chi $ is a dimensionless parameter equal to 400 for $\tau = 8$ years and 1000 for $\tau$ = $\infty$, which ensures that the average $S_{n}$ is identical to the average observed values.

For the SFT model parameters, our model grid was a subset of the grid described in Paper~1. For $\tau$, we only consider two values;
a decay time scale of 8 years, comparable to the cycle length and
supported by several studies (see Paper~1 and references therein), and a decay time scale of $\infty$, so no decay term. 
The meridional flow amplitude was set to $u_0=11$ \mm{m s$^{-1}$}, and the initial amplitudes of the inflow were $v_{00}= -500$ and $-750$ \mm{cm s$^{-1}$}, across various parameter combinations as detailed in Tables~\ref{table:inflows_M} and ~\ref{table:inflows_M_750}, respectively. \rev{The parameters were constrained in the range allowed by observations, and within this range, a limited mapping of the parameter space was performed.} These combinations correspond to two distinct initial inflow amplitudes, as suggested by \cite{Nagy+:inflow}. This approach allows us to demonstrate the impact of surface inflows with higher initial amplitudes.

\begin{table}
\caption{Set of parameters considered for inflows with the Surface inflows {whose amplitudes vary within a cycle depending on the magnetic activity} and its deviations $|\Delta D_n|$  of the nonlinear models from the linear case measured at a certain SSN value. The surface inflow initial amplitude $v_{00}=-500$ \mm{cm s$^{-1}$}.}
\label{table:inflows_M}
\begin{tabular}{lllllll}    
\hline  \\  
	Case & $u_{0}$ & $\eta$ & $\tau$ & $dev$ & $\devLQ$ & $dev$ \\
  & [\mm{m s$^{-1}$}] & [\mm{km$^{2}$s$^{-1}$}] & [yr] & \tiny LQ+inflows& & {\tiny inflows}  \\
\\
\hline
a&11&250&8&2.410&1.614&1.149\\
b&11&350&8&2.874&1.429&1.722\\
c&11&450&8&3.087&1.083&2.290\\
d&11&550&8&3.032&0.729&2.337\\
e&11&650&8&2.956&0.604&2.460\\
f&11&250&$\infty$&2.263&1.422&1.051\\
g&11&350&$\infty$&2.471&1.334&1.389\\
h&11&450&$\infty$&2.372&1.077&1.469\\
i&11&550&$\infty$&2.1621&0.763&1.445\\
j&11&650&$\infty$&2.040&0.637&1.331\\
\hline
\end{tabular}
\end{table}

\begin{table}
\caption{Same as Table~\ref{table:inflows_M}, with the initial surface inflow amplitude $v_{00}=-750 $ \mm{cm s$^{-1}$}.}
\label{table:inflows_M_750}
\begin{tabular}{lllllll}    
\hline  \\  
	Case & $u_{0}$ & $\eta$ & $\tau$ & $dev$  & $\devLQ$ & $dev$ \\
  & [\mm{m s$^{-1}$}] & [\mm{km$^{2}$s$^{-1}$}] & [yr] & \tiny LQ+inflows& & {\tiny inflows}  \\
\\
\hline

k&11&250&8&2.024&1.614&0.996\\
l&11&350&8&2.785&1.429&1.835\\
m&11&450&8&3.167&1.083&2.413\\
n&11&550&8&3.14&0.729&2.503\\
o&11&650&8&3.118&0.604&2.731\\
p&11&250&$\infty$&2.173&1.422&1.062\\
q&11&350&$\infty$&2.557&1.334&1.536\\
r&11&450&$\infty$&2.571&1.077&1.689\\
s&11&550&$\infty$&2.409&0.763&1.682\\
t&11&650&$\infty$&2.381&0.637&1.787\\

\hline
\end{tabular}
\end{table}

\section{Results}
\label{sect:SFTresults}

As the migrating flow perturbation represents a strong non-uniform latitudinal variation in the divergence of the flow, the problem at hand does not lend itself to the simplified analytic treatment applied in \cite{paper2}. Hence, the analytic fits to the results of our numerical exploration of the problem are simply based on a generalization of the results obtained by \cite{paper2}.

To investigate the effect of introducing surface inflows into the model, we first look into the impact of the surface inflows on the polar field for different setups (inflows, LQTQ, LQTQ+inflows) and compare it with an average cycle polar field. After that, we studied the model deviations in the net contribution of a cycle to the dipole moment $|\Delta D_{n}|$ for the three nonlinear scenarios (inflows, LQ and inflows+LQ) from the linear case (no LQ+ no inflows). Finally, we check the domination of LQ over TQ in the presence of surface inflows.

\subsection{Polar \mm{F}ield for \mm{Ten} Solar Cycles}

The polar field reverses by the end of every solar cycle, a process suggested by \cite{babcock1955sun} due to the magnetic flux transport on the solar surface. Correlation between the polar field at cycle minimum and the next cycle amplitude maximum confirmed by several studies \citep{wang2009understanding,hathaway2016predicting}, this makes the polar field a precursor of cycle strength \citep{Petrovay:LRSP2}.

Using the 1D SFT model described earlier, we look into the polar field for 10 solar cycles, considering the two nonlinear mechanisms, Latitude and tilt quenching (LQ and TQ), in addition to the surface inflows as a possible nonlinearity. First, we simulate an average solar cycle as in paper~1, adding both LQ and TQ to the model as in paper~2 and, finally, including the surface inflows. The model parameters considered \mm{in this article} are based on the optimisation results in paper~1. For meridional flow amplitude, $u_0=11$ \mm{m s$^{-1}$}, diffusivity $\eta$ was set to 250 \mm{km$^{2}$s$^{-1}$} and the decay time scale for $\tau= 8$ yr. For latitude quenching (LQ), the $b_{lat}$ parameter was set to 2.4, and for tilt quenching (TQ), the $b_{joy}$ parameter was set to 0.15.

\begin{figure}    
\centerline{\includegraphics[width=0.5\textwidth,clip=]{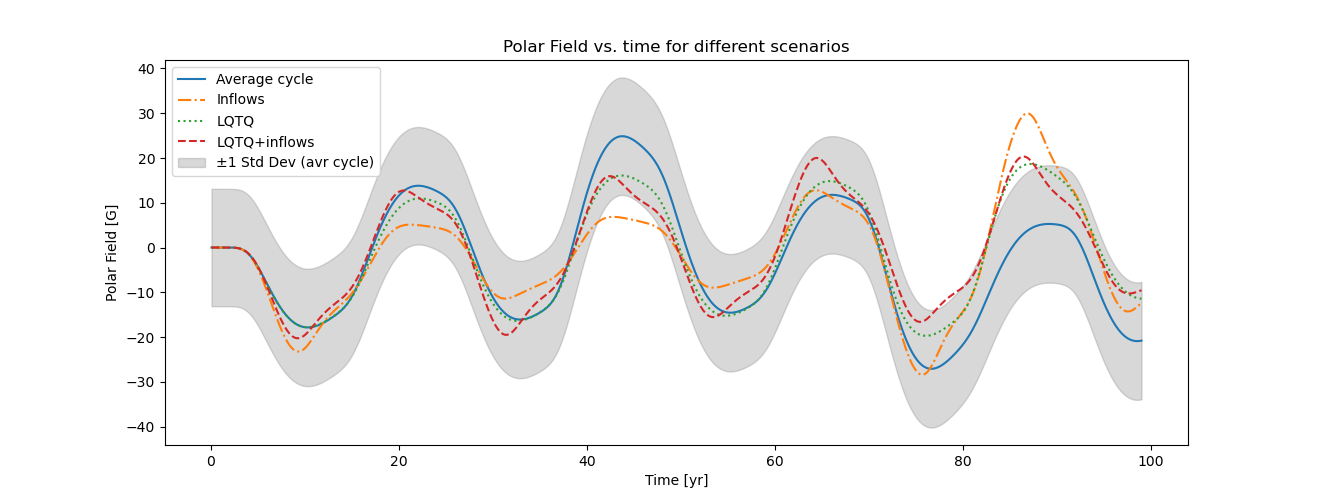}}
\small
    \caption{Polar field [G] $\it{vs.}$ Time [yr] for 10 average solar cycles (solid blue), including inflows (dash-dotted orange), latitude and tilt quenching (dotted green) and all of the three mechanisms to the average cycle (dashed red). A $\pm1\sigma$ of the average cycle is shown as the shaded area. }
    \label{fig:polar}
\end{figure}

The polar field values for different scenarios were calculated, firstly, polar field values for an average solar cycle, no quenching mechanisms and no inflows (average cycle), secondly, the average solar cycle with inflows (inflows), a third scenario to include both LQ and TQ with the average cycle polar field (LQTQ), and lastly to consider all the mechanisms together (LQTQ+inflows). \mm{Figure}~\ref{fig:polar} shows the polar field for 10 solar cycles for these different scenarios with time. A 1 Standard deviation ($\sigma$) is shown for the average cycle, the other three scenarios were nearly laying in between $\pm \sigma$, except for the last cycle which shows a slight deviation.

\begin{figure}    
\centerline{\includegraphics[width=0.5\textwidth,clip=]{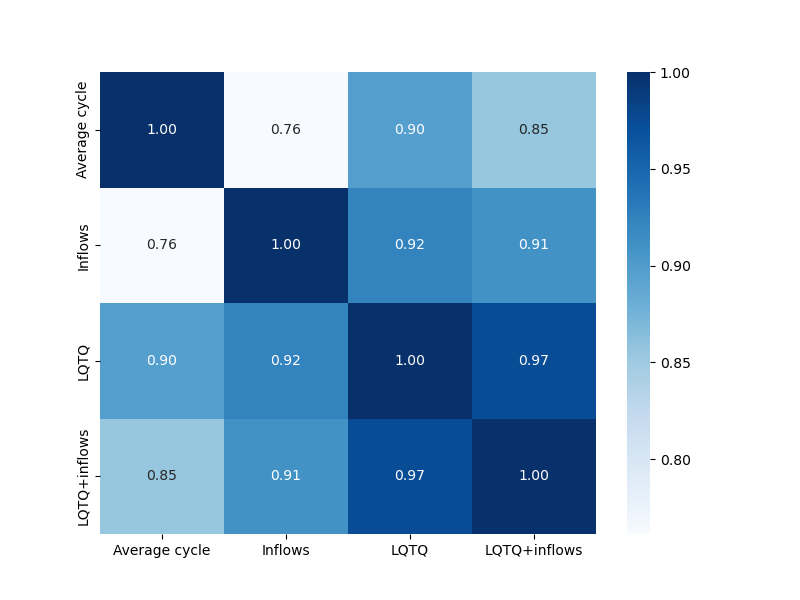}}
\small
        \caption{Correlation matrix illustrating the relationships between different cases. The matrix highlights the degree of correlation between parameters of the average cycle, inflows, latitude and tilt quenching (LQTQ), and combined (LQTQ+Inflows) scenarios. Higher values indicate stronger correlations, while lower values indicate weaker correlations.}

    \label{fig:heatmap}
\end{figure}

The correlation of the different scenarios with the average solar cycle is shown in \mm{Figure}~\ref{fig:heatmap}, it's important to say that the LQTQ scenario correlates with the average cycle by 0.9 factor, however, including the inflows to the model in addition to LQ and TQ reduces the correlation factor to 0.85.

\begin{figure}    
\centerline{\includegraphics[width=0.5\textwidth,clip=]{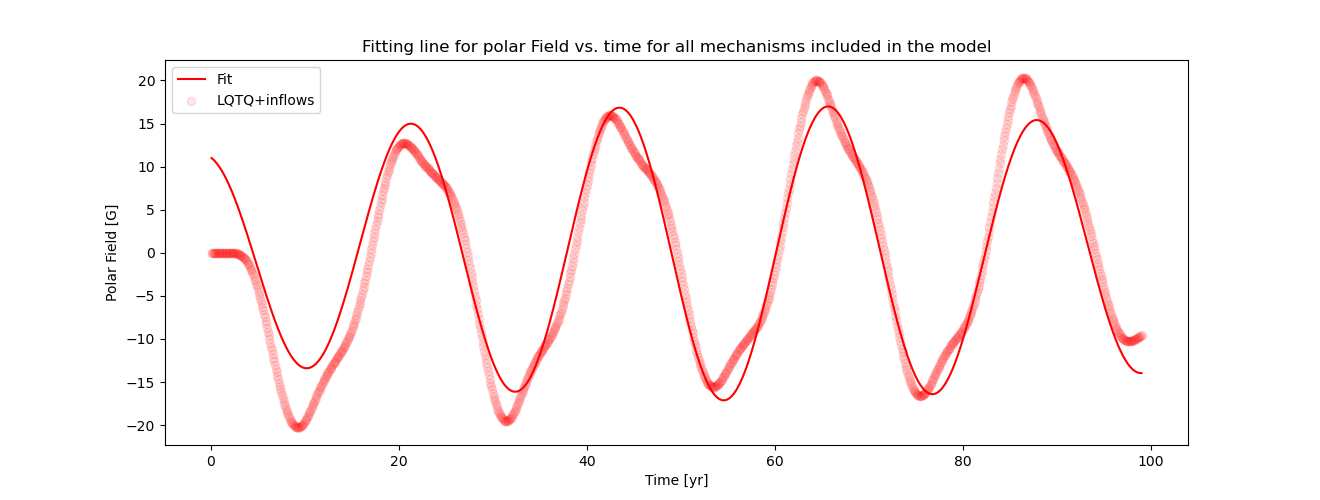}}
\small
    \caption{Fitting line for polar field vs. time for all mechanisms (LQTQ + inflows) case included in the model for 10 solar cycles, fitting parameters listed in Table~\ref{tab:fit_parm}}
    \label{fig:fit}
\end{figure}

The polar field for the average solar cycle was fitted to a harmonic function of the form:
\begin{equation}
    f(t)=(a_0 + a_1t+a_2t^{2})\sin{(bt+c)}
    \label{eq:fit1}
\end{equation}
where $a_0$, $a_1$, $a_2$, $b$ and $c$ are fitting parameters. The choice of this form was selected based on the varying amplitudes of the polar field through the 10 cycles in addition to its periodic nature. As an example, \mm{Figure}~\ref{fig:fit} shows the case where all of the three mechanisms considered (LQTQ+inflows) in the model, the polar field values were fitted to the form in \mm{Equation}~\ref{eq:fit1}, the complete list of the fitting parameters best values for each scenario given in Table~\ref{tab:fit_parm}.

 \begin{table}[ht!]
     \caption{Fitting parameters of polar field vs. time for different cases}
     \begin{tabular}{l|c|c|c|c|c}
         Case & $a_0$ & $a_1$& $a_2$& $b$ & $c$ \\
         \hline
         Average cycle & -8.986& -0.407&0.004&0.284&-1.54 \\
         LQ &-6.404& -0.378&0.003&0.284&-1.543 \\
         TQ &-10.583& -0.168&0.002&0.281&-1.417\\
         Inflows &-14.592& 0.282&-0.004&0.28&-1.145\\
         LQTQ &-9.296& -0.274&0.002&0.283&-1.453\\
         LQTQ+inflows &-11.587& -0.197&0.002&0.282&-1.27\\

         \hline
     \end{tabular}
     \label{tab:fit_parm}
 \end{table}

To compare each scenario with the average cycle, we calculate the percentage error between the fitting parameters of each scenario and the average cycle. Table~\ref{tab:percent_error} shows the percentage error for each parameter in \mm{Equation}~\ref{eq:fit1}. The "Inflows" scenario consistently shows the largest percentage errors across most parameters, particularly for \(a_1\) (169.29\%), \(a_2\) (200.00\%), and \(a_0\) (62.38\%), this proves that the inflows significantly impact the parameters compared to the other scenarios. The parameter \(b\) remains relatively stable across all scenarios, with the percentage errors being minimal (ranging between \mm{0.00\,--\,1.41\%}),  which suggests that \(b\) is less sensitive to the variations in the different cases studied. All scenarios show notable percentage errors in \(a_0\), but "Inflows" shows the largest positive percentage error; this parameter seems to be highly influenced by the inclusion of inflows. The combined cases (LQTQ and LQTQ+Inflows) generally show moderate percentage errors from the average cycle values, this suggests that the combined effects of Latitude and tilt quenching, as well as inflows, lead to a more balanced impact on the parameters. There is significant variation in parameter \(c\), especially for "Inflows" (25.65\%), indicating that the inflow case has a substantial impact on this parameter, differentiating it from other scenarios.

\begin{table}[ht!]
    \caption{Percentage error of fitting parameters for different cases compared to the average cycle.}
    \begin{tabular}{l|c|c|c|c|c}
        Case & $\Delta a_0 (\%)$ & $\Delta a_1 (\%)$ & $\Delta a_2 (\%)$ & $\Delta b (\%)$ & $\Delta c (\%)$ \\
        \hline
        LQ & 28.74 & 7.12 & 25.00 & 0.00 & 0.19 \\
        TQ & 17.77 & 58.72 & 50.00 & 1.06 & 7.99 \\
        Inflows & 62.38 & 169.29 & 200.00 & 1.41 & 25.65 \\
        LQTQ & 3.45 & 32.68 & 50.00 & 0.35 & 5.65 \\
        LQTQ+Inflows & 28.94 & 51.60 & 50.00 & 0.70 & 17.53 \\
        \hline
    \end{tabular}
    \label{tab:percent_error}
\end{table}

\subsection{Surface \mm{I}nflows as a \mm{N}onlinearity in the SFT \mm{M}odel}
\label{effect_quen}

The surface inflows can be represented by adding perturbations to the meridional flow profile. \cite{Nagy+:inflow} utilised the model framework proposed by \cite{jiang2010effect}, introducing modifications to the width and speed of the inflow to ensure their proportionality to the emerging magnetic flux. However, \mm{in this article}, we consider this profile to be dependent on the cycle amplitude, as described in the previous section. This modification transformed the inflows into an authentic nonlinear magnetic response mechanism rather than imposing predetermined latitudinal profiles and inflow speeds.  

\begin{figure}    
\centerline{\includegraphics[width=0.5\textwidth,clip=]{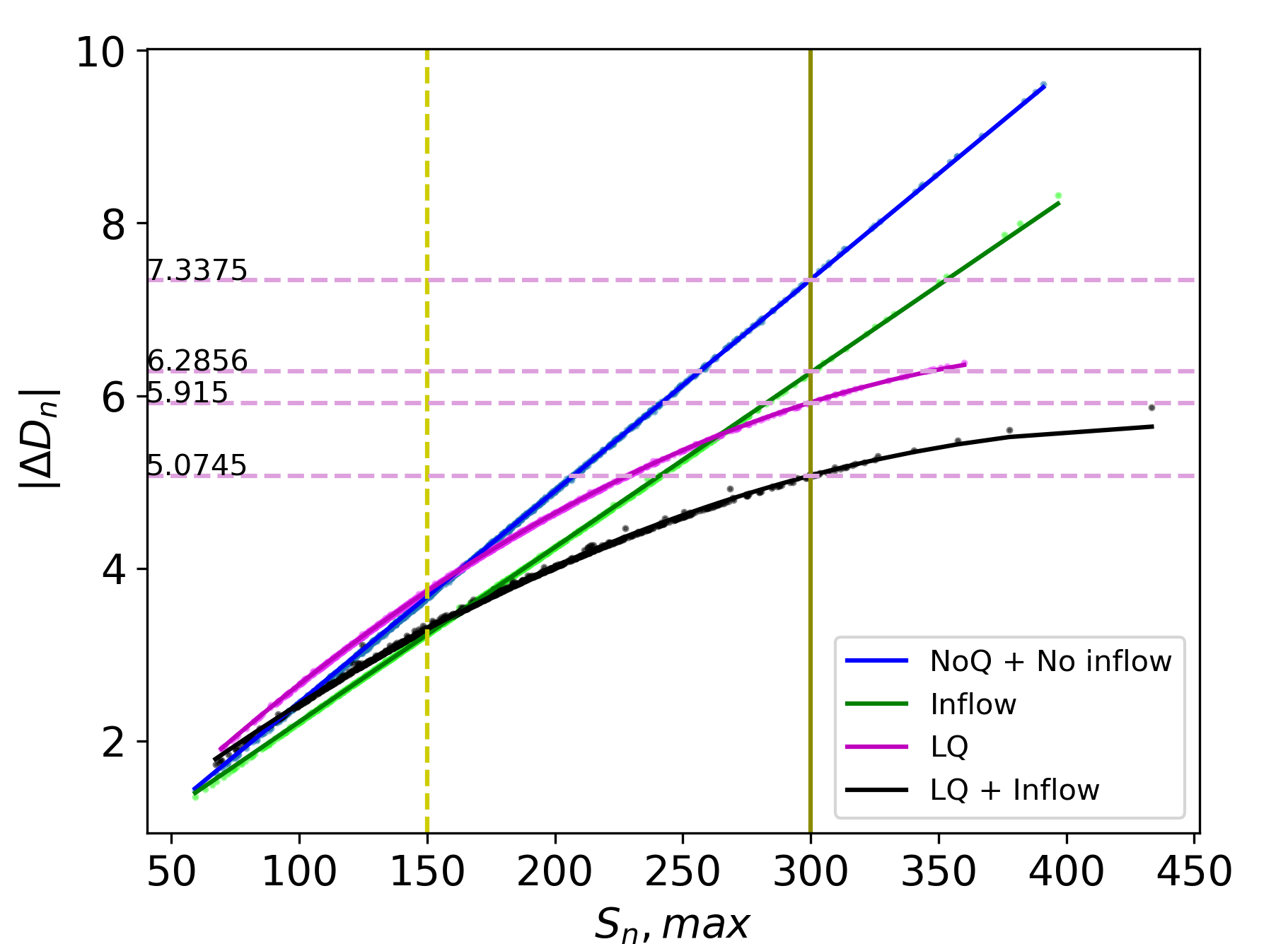}}
\small
        \caption{Net contribution of a cycle to the solar dipolar moment vs. cycle amplitude for the different cases to the linear case with no quenching. The parameter values were $u_0=11\,$m$/$s, $\eta=250\,$\mm{km$^{2}$s$^{-1}$} and $\tau = \infty $ years. inflows are implemented with an initial amplitude $v_00=-500$ \mm{cm s$^{-1}$}. The dashed and solid vertical lines show the average and twice average cycle amplitude values respectively, and the horizontal dashed lines show where the twice average vertical line intersects with each case.}
\label{fig:quad_fit_M}
\end{figure}

We aim to obtain a statistically meaningful sample of solar cycles by solving the azimuthally averaged SFT equation, Equation~\ref{eq:transp}, for 1000 cycles, by varying the parameters $\eta$, $\tau$ and $v_0$ as listed in Tables~\ref{table:inflows_M} and ~\ref{table:inflows_M_750}. As an illustration, \mm{Figure}~\ref{fig:quad_fit_M} shows the net contribution of a solar cycle to the solar dipole moment plotted against the cycle amplitude. This comparison was performed in the presence and absence of surface inflows.  Defining $r= e^{-11/\tau}$ (with $\tau$ in years), the net contribution of the cycle $n$ to the change in the dipolar moment is given by $\Delta D_n = D_{n+1}-r D_n$, where $D_n$ represent the dipole moment for the cycle $n$ at cycle minimum as defined in paper~2. 

TQ was not considered here because it might overestimate the effects of nonlinearities if both tilt quenching and inflows are incorporated into the surface flux transport process. The decision to exclude TQ and concentrate on LQ and inflows was motivated by our hypothesis that inflows may be responsible for tilt quenching. Inflows could play a significant role in the tilt-quenching process; larger tilt angles indicate that stronger inflows are required to obtain regular polar field reversals in SFT models, which brings new constraints on the amplitudes of inflow \citep{jiao2021sunspot}. \cite{martin2016surface} investigated the role of the converging flows towards a BMR in its evolution and their impact on the axial dipolar field, they found that the latitudinal separation of the polarities of the bipolar region is limited by the inflows, resulting in a reduction of the axial dipole moment of the BMR, and hence, lowers the contribution of the emerged BMR to the axial dipole moment. Therefore, to avoid the potential overestimation of nonlinear effects by incorporating both TQ and inflows into the SFT process, we examined the effects of LQ and inflows while excluding TQ.

Deviations from the linear case (no inflows + no LQ) were calculated at an arbitrarily selected value of $S_{n}$ (approximately twice the mean), as indicated by the solid vertical line in this figure. Table~\ref{table:inflows_M} shows the deviations of the inflows, LQ, and inflows+LQ from the linear case (no inflows + no LQ), for example, in \mm{Figure}~\ref{fig:quad_fit_M}, $dev_{inflows}$ = 7.335-6.285 = 1.051.

The average $S_{max}$ was set to 150 based on the literature. Therefore, for each parameter combination, {$S_{n},max$} was scaled using $\chi$ to obtain values comparable to the observed sunspot number SSN. For various cases (a-t), it is evident that these deviations are significantly influenced by the selection of $u_{0}$, $\eta$, and $\tau$. The linear case (no inflows + no LQ) was fitted with a linear function, whereas the other three cases were fitted with quadratic polynomials following the results in paper~2. 

Including inflows in the model resulted in a decreased net contribution to the solar dipole moment in all cases, as shown in \mm{Figure}~\ref{fig:quad_fit_M}. For both linear and LQ cases, having inflows in the model diminished the overall cycle contribution to the solar dipole moment, as indicated by the $dev_{\text{inflows}}$ and $dev_{\text{LQ+inflows}}$ in Tables ~\ref{table:inflows_M} and ~\ref{table:inflows_M_750}.

\begin{figure}    
\centerline{\includegraphics[width=0.5\textwidth,clip=]{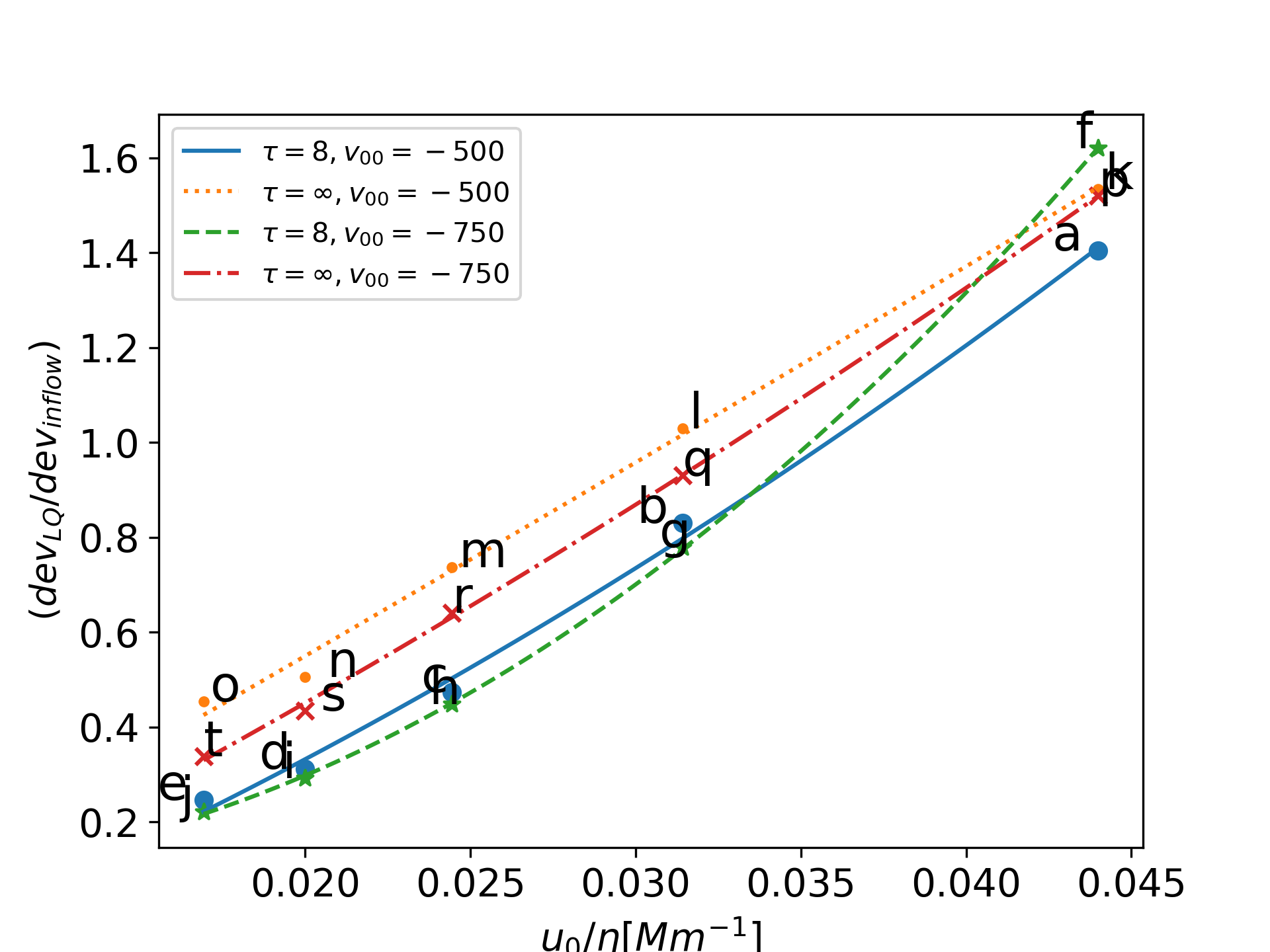}}
\small
        \caption{The relative importance of LQ vs. inflows plotted against the $u_{0}/\eta$ ratio. Separate third-degree polynomial fits for the case $\tau = 8$ (solid blue and dashed green) and $\tau = \infty$ (dotted orange and dash-dotted red) are shown for the different parameter combinations listed in Tables~\ref{table:inflows_M} and \ref{table:inflows_M_750}.}
\label{fig:inflow_M}
\end{figure}

Figure~\ref{fig:inflow_M} shows the relative importance of LQ vs. inflows for different parameter combinations plotted against $u_0/\eta$ and fitted to a quadratic polynomial of the form:
\begin{equation}
\label{eq:fit}
    dev_{\text{LQ}}/dev_{\text{inflows}}=c_{1} (u_{0}/\eta)^2 + c_{2}(u_{0}/\eta)+ c_{3}
\end{equation}
where $c_{1}$, $c_{2}$, and $c_{3}$ are the fitting parameters that depend on $\tau$. Note that the relative importance of the LQ vs. inflows for $\tau$ = 8 is always higher than that for $\tau = \infty$; that is, with no decay term, inflows will be less significant. Using two initial inflow amplitudes $v_{00}$, introducing the surface inflows with $v_{00}=-750$ \mm{cm s$^{-1}$} gave lower relative importance of LQ vs. inflows than using $v_{00}=-500$ \mm{cm s$^{-1}$} for both $\tau= 8$ and $\infty$. To assess whether the fitting parameters $c_1$, $c_2$ and $c_3$ differed significantly between the two inflows amplitudes $v_{00}=-500$ \mm{cm s$^{-1}$} and $v_{00}=-750$ \mm{cm s$^{-1}$}, we conducted paired t-tests for each parameter. The parameters were analyzed under the two values of $\tau =$ 8 years and $\infty$. Table~\ref{tab:ttest_results} show the t-statistics and p-values, these results indicate that there are no significant differences in the fitting parameters \(c1\), \(c2\), and \(c3\) between the two initial inflows amplitudes for both \(\tau = 8\) and \(\tau = \infty\).
 This can be interpreted as increasing the initial inflow amplitudes produced slightly significant inflows.

\begin{table}[ht!]
\caption{Paired t-test results for fitting parameters \(c1\), \(c2\), and \(c3\) comparing ($v_{00}$ = -500\ \mm{cm s$^{-1}$}) and ($v_{00}$ = -750\ \mm{cm s$^{-1}$}) for both $\tau = 8$ [yr] and $\infty$.}
\begin{tabular}{lcc}
\hline
Parameter & t-statistic & p-value\\
\hline
$c_1$  & -1.60 & 0.35 \\
$c_2$  & 1.52 & 0.37 \\
$c_3$  & -0.92 & 0.53 \\
\hline
\end{tabular}
\label{tab:ttest_results}
\end{table}

The dynamo effectivity range $\lambda_{R}$ enabled visualization of the inherent connection between the nonlinearity ratios related to the inflows and $\lambda_{R}$. {where $\lambda_{R}$ defined by the empirical fit found by \citet{Petrovay+:algebraic1}:}

\begin{equation}
\label{eq:lambdaR}
    \lambda_{R,fit}= g^{1/2}(x=\eta/R^{2}\Delta_{u})\lambda_{R,limit}
\end{equation}
with 

\begin{eqnarray}
\label{eq:lambdaR1}
    g(x)&=& (m_1x+c_1)\{1-tanh[(x-c_{0})/w]\} \nonumber\\
    &&+(m_2x+c_2)\{tanh[(x-c_{0})/w]\}
\end{eqnarray}

and $m_1=0.0004$, $m_2=-0.0008$, $c_0=200$, $c_1=1.18$, $c_2=1.42$. Note that $\Delta_{u}$ is the divergence of the meridional flow at the equator.
\begin{equation}
\label{eq:deltau}
    \Delta_{u}= \frac{1}{R}{\frac{du}{d\lambda}}|_{\lambda=0}
\end{equation}

\begin{figure}    
\centerline{\includegraphics[width=0.5\textwidth,clip=]{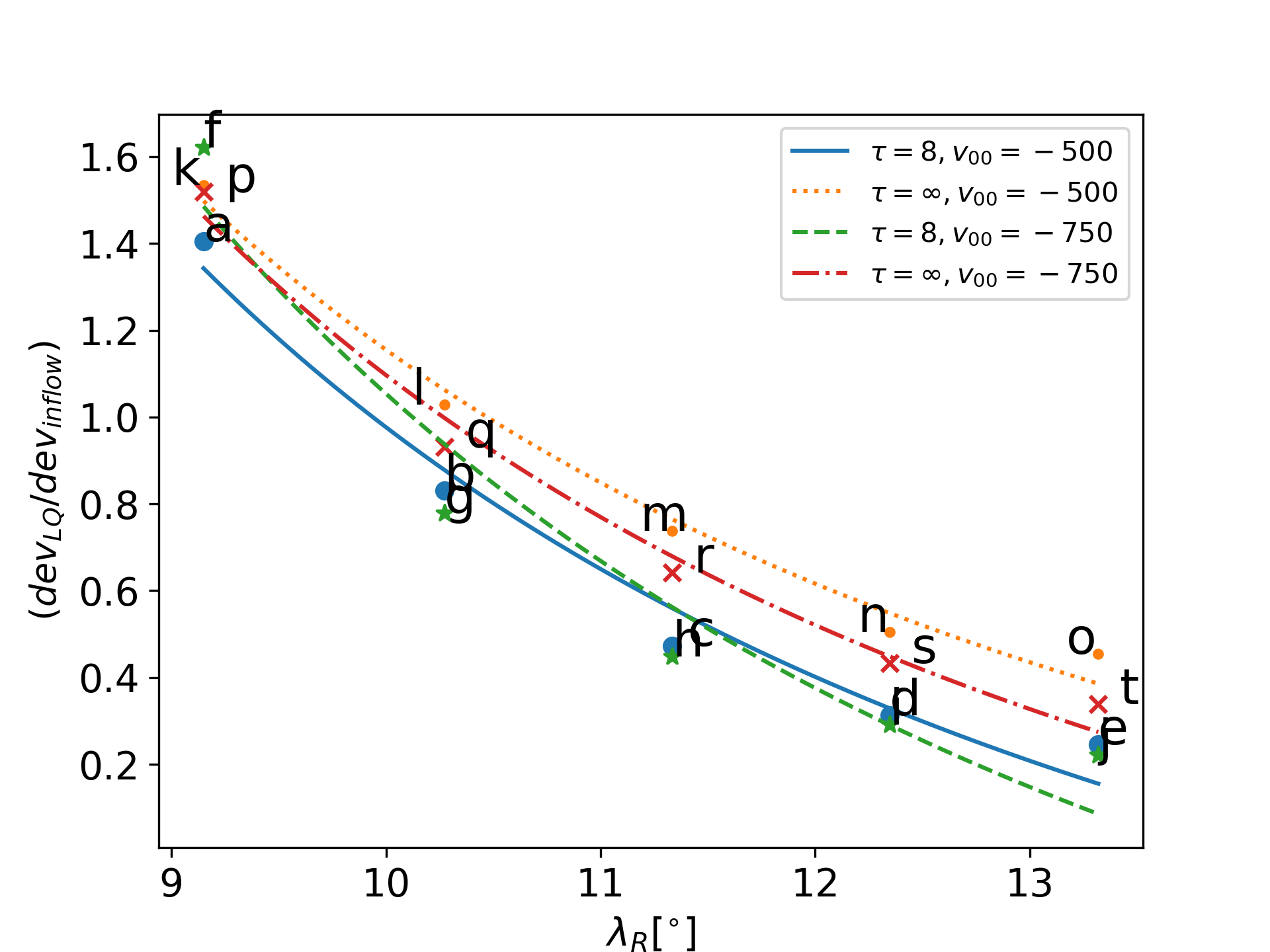}}
\small
        \caption{The relative importance of LQ vs. inflows plotted against the dynamo effectivity range, $\lambda_R$, for the parameter combinations listed in Tables~\ref{table:inflows_M} and \ref{table:inflows_M_750}. Distinct fit to \mm{Equation}~\ref{eq:fit2} are shown for $\tau=8$ (solid blue and dashed green ) and $\tau=\infty$ (dotted orange and dash-dotted red), respectively.}
\label{fig:lambda_r_M}
\end{figure}

\mm{Figure}~\ref{fig:lambda_r_M} show the relative importance of LQ vs. inflows to $\lambda_R$ fitting lines for both $\tau=8$ and $\tau=\infty$ and two initial inflows amplitudes, the fitting lines were structured according to \mm{Equation}~\ref{eq:fit2}, for a higher amplitude of the inflow $v_{00}=-750$ \mm{cm s$^{-1}$}, the relative importance of LQ vs. inflows is less significant for $\tau =$$\infty$. However, for $\tau = 8$ yr, the relative importance of LQ vs. inflows is slightly higher for low $\lambda_R$.

As a consequence of the inflows, the net flow will become convergent within a belt of half-width $\sim\delta\lambda_0$, centred on $\lambda_0$. In the absence of diffusion, the fraction $f_0$  of the flux in the source that falls within the convergent belt would be advected to $\lambda_0$ and cancelled. Diffusion allows a fraction $f_1$ of this flux to escape from the convergence belt; therefore, in the final accounting, a fraction $f=1-f_0(1-f_1)$ of the flux escapes cancellation and evolves under the effect of the large-scale meridional flow and diffusion, essentially unaffected by the inflows. As $f_1$ depends on the inflow speed, $f$ is modulated by the cycle amplitude. Therefore, the net effect of inflow modulation is formally reduced to the modulation of the source term in the SFT model, as in the case of tilt quenching. 
Therefore, we expect the effects of inflow modulation to be similar to those of the TQ. In particular, for cycle amplitudes varying in a smaller range, the resulting modulation of the source is expected to remain in the linear regime similar to the form of TQ treated in paper~1, where it was found that the relative importance of LQ and TQ will scale with the dynamo effectivity range as $C_1+C_2/\lambda_R^2$. This prompts us to fit the curves in \mm{Figure} 3 with the same fitting formulae,

\begin{equation}
\label{eq:fit2}
    dev_{\text{LQ}}/dev_{\text{inflows}}=c_{1} + c_{2}/(\lambda_R)^2 
\end{equation}
where $c_{1}$ and $c_{2}$ are fitting parameters.
The fit is almost perfect. Small deviations are due to numerical errors and may be attributed to the fact that cycle amplitudes vary over a rather wide range; hence, the assumption of linear dependence of modulation of the source on cycle amplitude is rather crude. 

\begin{figure}    
\centerline{\includegraphics[width=0.5\textwidth,clip=]{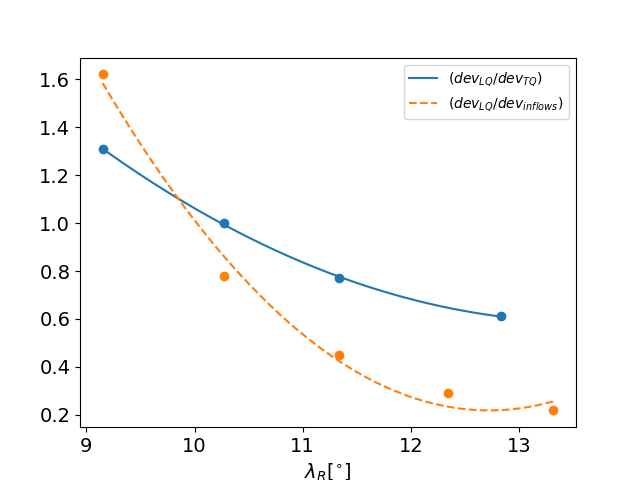}}
\small
        \caption{The relative importance of LQ vs. TQ from \citet{paper2} (solid blue) and LQ vs. inflows (dashed orange) plotted against the dynamo effectivity range $\lambda_R$, for $\tau=8$, $v_{00}=-750$ \mm{cm s$^{-1}$}, and parameter combinations as listed in Table~\ref{table:inflows_M_750}. Fit lines to a quadratic polynomial are shown as in \mm{Equation}~\ref{eq:fit}.}
\label{fig:comp_lambda_r}
\end{figure}
\begin{figure}    
\centerline{\includegraphics[width=0.5\textwidth,clip=]{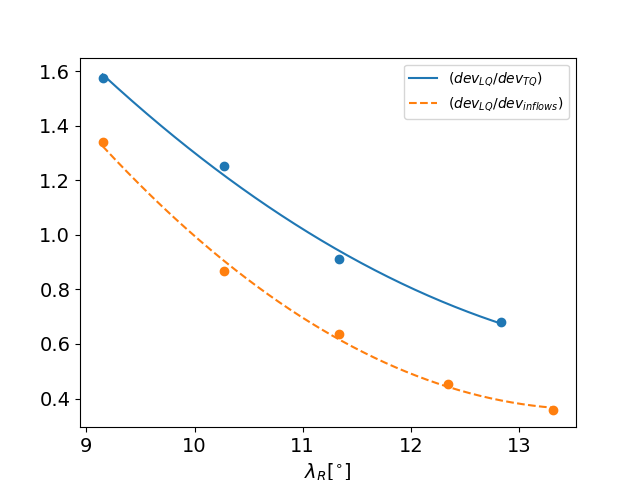}}
\small
        \caption{Same as \mm{Figure}~\ref{fig:comp_lambda_r} for $\tau=\infty$, $v_{00}=-750$ \mm{cm s$^{-1}$}, and parameter combinations as listed in Table~\ref{table:inflows_M_750}.}
\label{fig:comp_lambda_r_1000}
\end{figure}

Figure~\ref{fig:comp_lambda_r} and ~\ref{fig:comp_lambda_r_1000} show a comparison between the relative importance of TQ vs. LQ from paper~2 and the relative importance of inflows vs. LQ to $\lambda_R$ for $\tau=8$ years and $\infty$ respectively. The TQ and inflows exhibit similar behaviour. However, the relative importance of TQ vs. LQ is always larger than LQ vs. inflows; this supports our initial hypothesis that inflows may be responsible for tilt quenching, and hence, considering both will overestimate the effects of the nonlinearities.

\subsection{Domination of LQ over TQ when considering \mm{S}urface \mm{I}nflows}

To determine the domination of LQ over TQ we solve the azimuthally averaged SFT equation for 1000 cycles, by varying the parameters $\eta$, $\tau$ and considering no inflows in one case; $v_{00}=0$, and with inflows in the second case; $v_{00}=-500$ [\mm{cm s$^{-1}$}], as listed in  Table~\ref{table:dev}. The meridional flow amplitude $u_{0}$ was set as 11 [\mm{m s$^{-1}$}] for all cases.

\begin{table}[t]
    \caption{Deviations of TQ and LQ from the linear case (NoQ) for different parameter combinations for no inflows $v_{00}=0$ and with inflows $v_{00}=-500$ \mm{cm s$^{-1}$}.}
    \label{table:dev}
    \begin{tabular}{@{}cccccccc@{}}
    \hline  \\  
    \multicolumn{2}{c}{Case} &  \multicolumn{3}{c}{$v_{00}=0$} &  \multicolumn{3}{c}{$v_{00}=-500$ [\mm{cm s$^{-1}$}]} \\\hline
 $\eta$ & $\tau$ & $dev_{TQ}$ &$dev_{LQ}$ &$[dev_{TQ}-dev_{LQ}$] & $dev_{TQ}$ &$dev_{LQ}$ &$[dev_{TQ}-dev_{LQ}]$ \\
\hline
250&8&1.271&1.779&-0.508&0.911&1.051&-0.140\\
350&8&1.303&1.342&-0.039&1.181&1.148&0.033\\
450&8&1.447&1.197&0.250&1.260&1.029&0.231\\
550&8&1.586&0.940&0.647&1.511&1.178&0.333\\
650&8&1.593&0.845&0.749&1.552&0.962&0.589\\
250&$\infty$&0.979&1.431&-0.452&0.941&1.378&-0.437\\
350&$\infty$&1.099&1.284&-0.185&1.125&1.320&-0.196\\
450&$\infty$&0.938&0.966&-0.028&1.170&1.191&-0.021\\
550&$\infty$&1.121&0.719&0.403&1.203&1.099&0.104\\
650&$\infty$&1.053&0.604&0.449&1.313&0.947&0.367\\
\hline
\end{tabular}
\end{table}

To investigate the effect of introducing surface inflows on the domination of LQ over TQ, we studied the model deviations of the two nonlinear scenarios (TQ and LQ) from the linear case (NoQ) for the net contribution of a cycle to the dipole moment $|\Delta D_{n}|$ with and without inflows. Table~\ref{table:dev} shows these deviations and the differences of the deviations ($dev_{TQ}-dev_{LQ}$). The model was run for a range of values of $\eta$ $\in$ [250 - 650] km$^2$ s$^{-1}$, and $\tau $ $\in \{8, \infty \}$. 

\begin{figure}    
\centerline{\includegraphics[width=0.5\textwidth,clip=]{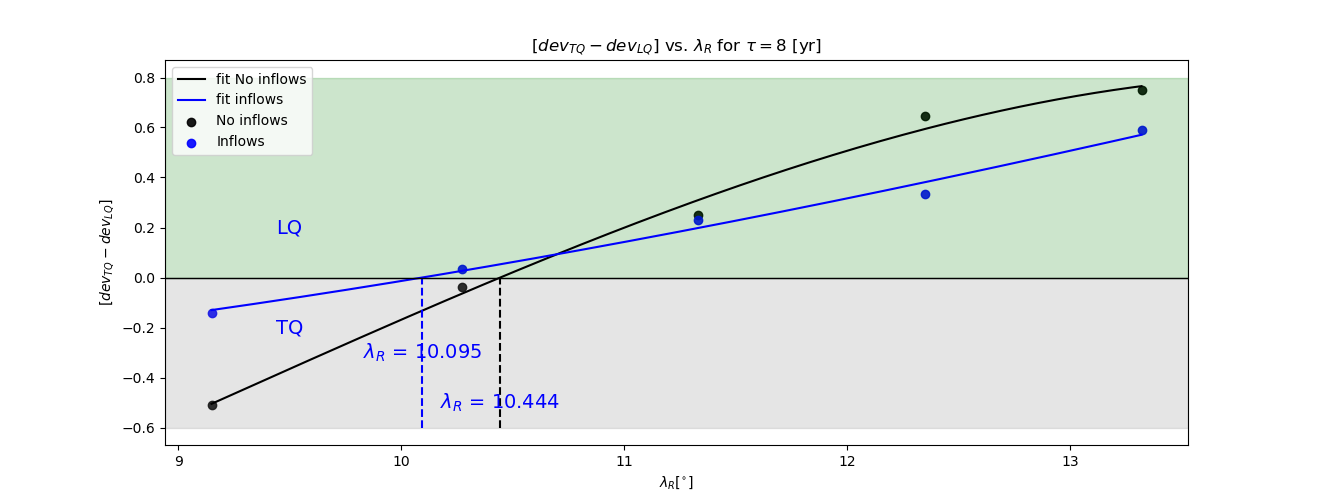}}
\small
    \caption{Difference of the deviations ($dev_{TQ}-dev_{LQ}$) plotted against $\lambda_R$ for $\tau = 8$ years, fitting lines are plotted of third-degree polynomial, the vertical lines show the points of transition from TQ to LQ.}
    \label{fig:diff8}
\end{figure}

\begin{figure}    
\centerline{\includegraphics[width=0.5\textwidth,clip=]{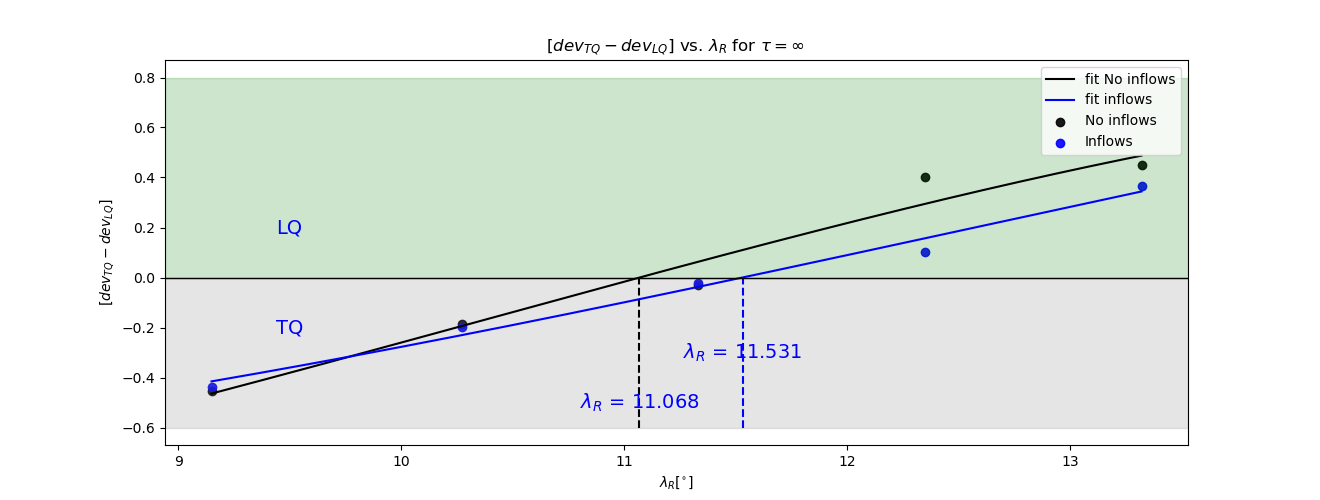}}
\small
    \caption{Same as \mm{Figure}~\ref{fig:diff8} for $\tau = \infty$. }
    \label{fig:diffinf}
\end{figure}

Initially, for both values of $\tau$, $dev_{TQ}$ is lower than $dev_{LQ}$, which means that the net contribution of a cycle to the solar dipole moment is higher for TQ and it dominates over LQ, by increasing the $\eta$ value the net contribution of a cycle to the solar dipole moment for TQ decreases gradually until the LQ dominates over TQ (at $\lambda_R =10.44 ^\circ$ for $\tau =8$ and $\lambda_R = 11.068 ^\circ$ for $\tau = \infty$). This is shown in Figures~\ref{fig:diff8} and \ref{fig:diffinf}; the vertical lines show the points of the transitions from TQ to LQ.

Adding inflows to the model reduced the net contribution of a cycle to the solar dipole moment and this sped up the domination of LQ over TQ for $\tau = 8$ years (at $\lambda_R = 10.095 ^\circ$), this can be attributed to the presence of inflows, having inflows in the SFT model reduces the flux transport toward the poles which will reduce the tilt angle and produce a lower net contribution to the dipole moment. This is not the case for $\tau = \infty$; introducing inflows to the model makes the transition from TQ to LQ slightly later in $\lambda_R = 11.531 ^\circ$. This can be interpreted as $\tau=\infty$; there is no decay term, the inflows contribute to the model, and less flux cancellation occurs on the surface; this will provide more flux to the active regions and, hence, increase the tilt angle, which makes TQ dominate a little bit longer over LQ. This supports our previous result for the decay term's importance and gives evidence for our initial hypothesis that inflows might be responsible for the tilt quenching.

\section{Discussion}
\label{sect:dis}

The presence of surface inflows in solar cycle modulation, represented by perturbations to the meridional flow profile, reduces the different cycle characteristics. \citet{cameron2012strengths} show that the magnetic flux corresponding to the axial dipole is reduced by applying strong inflows in an SFT model (\mm{Figure}~2 therein). \cite{martin2016surface} investigated the role of converging flows towards a BMR in its evolution and their impact on the axial dipolar field, they found that the latitudinal separation of the polarities of the bipolar region is limited by the inflows, resulting in a reduction of the axial dipole moment of the BMR, and hence, lowers the contribution of the emerged BMR to the axial dipole moment.

Adding inflows to the model produces a polar field within $\pm \sigma$ of the polar field of an average cycle.
The correlation coefficients between the average cycle polar field and the three mechanisms were relatively high, including inflows only in the model, which gave a 0.76 correlation coefficient, and combined with LQ and TQ, which showed a 0.85 correlation coefficient.
The polar field fitting parameters for considering inflows only in the model show the largest percentage of error, however, including LQ and TQ with the inflows showed a moderate percentage of error, which led to a more balanced impact on the parameters. This comes in agreement with the findings of \citet{martin2017inflows} and \citet{Nagy+:inflow} where they show that including inflows produces an alteration in the transportation of magnetic flux toward the pole and this alteration leads to a reduction in the cross-equatorial cancellations of BMRs and suppresses the effectiveness of the Babcock–Leighton process.

However, according to \citet{Nagy+:inflow}, a reduction of \mm{10\,--\,20\%} in the strength of the global dipole as it forms towards the end of a cycle compared to a scenario without inflows, this reduction was observed using a 2$\times$2D Babcock–Leighton solar cycle model. Our simulation produced close results to \citet{Nagy+:inflow} by comparing the net contribution of a cycle to the solar dipolar moment with and without inflows in the absence of LQ, 
as in \mm{Figure}~\ref{fig:quad_fit_M} for example, a \mm{12.6\,--\,22.8\%} reduction in case of $v_{00}=-500$ \mm{cm s$^{-1}$} and \mm{10.9\,--\,25.4\%} reduction for $v_{00}=-750 $ \mm{cm s$^{-1}$} using the SFT models and depending on the parameter combination, overall, when $\tau = \infty$ the reduction of the global dipole is less than the cases with $\tau = 8$ years.

Using the surface inflow profile influenced by the total emerged magnetic flux from the past cycle, \citet{Nagy+:inflow} adjusted the width and speed of the inflow to be directly related to the emerging magnetic flux, effectively transforming them into a valid nonlinear magnetic feedback mechanism.  Figure~\ref{fig:lambda_r_M} shows the relative importance of LQ vs. inflows plotted against the dynamo effectivity range, $\lambda_{R}$, for $\tau = 8$, the relative importance of LQ vs. inflows always dominated by the values for $\tau =\infty$, which was the same of the relative importance of LQ vs. TQ  (as illustrated in \mm{Figure}~4, paper~2). This indicates that surface inflows are strongly affected by the presence of a decay term regarding the initial inflow amplitude. As discussed earlier in \mm{Section}~\ref{sect:SFTresults}, the net flow will become convergent within a belt of half-width $\delta\lambda_0$ centred on $\lambda_0$, and for strong diffusion, the escaped fraction of the flux from the convergent belt $f_1$ will increase, which means less flux escape cancellation $f$. This can be seen in the $dev_{inflows}$ values, which decrease with increasingly strong radial diffusion. On the other hand, the evolved flux under the effect of large-scale meridional flow and diffusion increases the latitude quenching mechanism, which can explain the increase in $dev_{LQ}$ with stronger diffusion.

Figure~\ref{fig:lambda_r_M} shows the relative importance of LQ vs. inflows against $\lambda_R$; fits of the form $C_1+C_2/\lambda_R^n$ were attempted, with different integer values for $n$, where $n=2$ was found to produce a reasonably good representation of the results. This agrees with the results paper~2 for the relative importance of LQ vs. TQ, which show similar behaviour, suggesting that LQ is dominant over surface inflows in the absence of TQ. {This similarity is illustrated in Figs.~\ref{fig:comp_lambda_r} and ~\ref{fig:comp_lambda_r_1000}. However, the relative importance of TQ vs. LQ is always greater than the relative importance of LQ vs. inflows to $\lambda_R$, which implies that a fraction of the TQ nonlinearities can be attributed to the inflow effects and also supports our hypothesis that inflows might be responsible for TQ.}

The relative importance of LQ vs. inflows for the different sets of parameters (roughly between 0.2 and 1.6) was similar to the relative importance of LQ vs. TQ. However, the relative importance of LQ vs. TQ is between 0 and 2, with some negative values. {Most of the SFT and dynamo models studied in the recent literature {\citep{Bhowmik_Nandy_2018, Jiang+Cao, Lemerle2, Whitbread+:SFT},} were in the range of 1 \,-- \,2. } This can be explained by examining the amplitude of the surface inflow; having inflow amplitudes depending on the cycle amplitude means that we will have a lower net contribution to the dipole moment (Figure~\ref{fig:quad_fit_M}), which results in higher inflows deviations $(dev_{\text{inflows}}$), and the relative importance of LQ vs. inflows decrease.

{The nonlinear relationship between the relative importance of LQ vs. inflows to the dynamo effectivity range $\lambda_{R}$ is evident as illustrated in \mm{Figure}~\ref{fig:lambda_r_M} for various combinations of parameters. The fitted lines demonstrate an inverse proportionality between $\lambda_{R}$ and the relative importance of LQ vs. inflows, as modelled by \mm{Equation~\ref{eq:fit2}}. This underscores the nonlinear nature of surface inflows as a truly nonlinear mechanism.}

An analysis using a paired t-test on the fitting parameters for both values of $v_{00}$ showed no significant differences. This indicates that varying the initial amplitude of the inflow has only a minor effect on the importance of the inflows.

For lower values of $\lambda_R$, TQ dominates over LQ with and without including surface inflows, by increasing $\lambda_R$ gradually, a smooth transition for LQ to be dominating the TQ. This come in agreement with paper~2, where it was shown that when $\lambda_{R}$ is small, LQ exhibits a greater reduction in the end-of-cycle axial dipole than TQ and conversely, for large values of $\lambda_{R}$, TQ is more influential than LQ in diminishing the end-of-cycle dipole. However, including inflows makes this transition occur faster for $\tau = 8$ yr. However, for no decay term ($\tau = \infty$), TQ dominate a little bit longer over LQ.

\mo{We reiterate that the parameter study in the present work is somewhat limited. A much more comprehensive mapping of the parameter space for the time-independent flow and transport parameters was performed by \citet{paper1}. The observational constraints applied in that study were based on phase differences in the temporal variation of various measures of the solar magnetic field. An interesting further possibility to constrain the parameters was pointed out by Petrie (\citeyear{petrie2023,petrie2024global}), who suggested that characteristic shapes of the flux patterns originating from decaying active regions at higher latitudes may serve as sensitive tests of the transport parameters. Exploiting these constraints in a 2D flux transport model is one potential direction for a future extension of this work.}

\section{Conclusion}
\label{sect:concl}

In this study, we used the 1D surface flux transport model to study the effect of nonlinear surface inflows into activity belts in solar cycle modulation. We used a modified meridional flow perturbation profile presented by\citet{nagy2017effect} to represent the surface inflows whose amplitudes vary within a cycle depending on the magnetic activity.

Our results confirmed that introducing surface inflows into the solar dynamo will provide a polar field within an average cycle polar field of $\pm 1\sigma$, and including LQ and TQ together with inflows will give a more balanced impact on the polar field. Results show that having inflows will decrease the strength of the global dipole moment in the presence or absence of latitude quenching, and for different allowed parameter combinations. Our model yielded results that closely resemble those of the 2$\times$2D dynamo model. A reduction of \mm{10\,--\,25\%} was observed depending on the chosen model parameters.

Including inflows into the model shows that the relative importance of LQ vs. inflows is inversely correlated with the chosen parameters ($u_0, \eta$, and $\tau$) and hence, on the dynamo effectivity range ($\lambda_{R}$), with lower values depending on the initial amplitude of the surface inflow.

When there is no decay term ($\tau$=$\infty$), the introduction of the inflows into the model results in a less significant net contribution to the dipole moment. In other words, a decay term is essential for the model.

For the used inflow profile, the anticipated modulation of the source was expected to persist within the linear regions, similar to the TQ case. However, relying on fitting to determine the significance of LQ vs. inflows suggests that the assumption of a linear dependence of source modulation on cycle amplitude is somewhat simplistic. This shows a possible nonlinear relationship between the surface inflows and the solar dipole moment, suggesting a potential nonlinear mechanism contributing to the saturation of the global dynamo within the Babcock-Leighton framework.

TQ dominates over LQ for low $\lambda_R$, after that a transition for LQ gradually dominates TQ, adding inflows gives an earlier transition of domination from TQ to LQ. However, for no decay term, the transition will occur a little bit later.

A plausible further extension of this work would consider solar cycle scale variation of the order of 2 \mm{m s$^{-1}$} as the third near-surface meridional flow component, in addition to the constant baseline flow profile and the variations due to inflows around active regions.



\begin{ethics}
\begin{conflict}
The authors declare that they have no conflicts of interest.
\end{conflict}
\end{ethics}

\begin{authorcontribution}
M.T. led the investigation and drafted the majority of the manuscript, K.P. played a key role in designing the experiment and enriching the discussion, and A.O. meticulously reviewed the manuscript to refine its language and clarity.
\end{authorcontribution}

\begin{fundinginformation}
\mo{This research was supported partially by the Hungarian National Research, Development and Innovation Fund (grant numbers NKFI FK-128548 and TKP2021-NKTA-64), as well as by the European Union's  Horizon 2020 research and innovation programme (grant no.~955620)}.
\end{fundinginformation}

\begin{dataavailability}
No data have been produced during this work.
\end{dataavailability}

\bibliographystyle{spr-mp-sola}
\bibliography{sola_bibliography_example}  

\end{document}